\documentclass[%
reprint,
amsmath,amssymb,
aps,
prd,
nofootinbib, 
]{revtex4-2}

\usepackage{graphicx}
\usepackage{dcolumn}
\usepackage{bm}
\usepackage{color}
\usepackage{lipsum}
\usepackage{hyperref}

\usepackage{array, multirow} 
\usepackage{hhline}

\usepackage{relsize,exscale}
\usepackage{array}
\usepackage{ctable}
\usepackage{cellspace} 
\newcolumntype{?}{!{\vrule width 0.12em}}

\usepackage[normalem]{ulem}

\usepackage{relsize,exscale}
\usepackage{array}
\usepackage{ctable}
\usepackage{cellspace} 
\usepackage{nicefrac}
\newcolumntype{?}{!{\vrule width 0.12em}}

\begin{document}

\title{Circular reasoning: Solving the Hubble tension with a non-$\pi$ value of $\pi$}

\author{Jonas {El~Gammal}$^1$}
\email{jonas.e.elgammal@uis.no}
\author{Sven Günther$^2$}
\email{sven.guenther@rwth-aachen.de}
\author{Emil~Brinch Holm$^3$}
\email{ebholm@phys.au.dk}
\author{Andreas Nygaard$^3$}
\email{andreas@phys.au.dk}

\affiliation{
        \vspace{0.2cm}
        $^1$Department of Mathematics and Physics, University of Stavanger, NO-4036 Stavanger, Norway
}
\affiliation{
	\vspace{0.2cm}
	$^2$ Institute for Theoretical Particle Physics and Cosmology (TTK), RWTH Aachen University, DE-52074 Aachen, Germany
}
\affiliation{
	\vspace{0.2cm}
	$^3$Department of Physics and Astronomy, Aarhus University, DK-8000 Aarhus C, Denmark 
        \vspace{0.2cm}
}

\date{April 1, 2024}

\begin{abstract}
Recently, cosmology has seen a surge in alternative models that purport to solve the discrepancy between the values of the Hubble constant $H_0$ as measured by cosmological microwave background anisotropies and local supernovae, respectively. In particular, many of the most successful approaches have involved varying fundamental constants, such as an alternative value of the fine structure constant and time-varying values of the electron mass, the latter of which showed particular promise as the strongest candidate in several earlier studies. Inspired by these approaches, in this paper, we investigate a cosmological model where the value of the geometric constant $\pi$ is taken to be a free model parameter. Using the latest CMB data from \textit{Planck} as well as baryon-acoustic oscillation data, we constrain the parameters of the model and find a strong correlation between $\pi$ and $H_0$, with the final constraint $H_0 = 71.3 \pm 1.1 \ \mathrm{ km/s/Mpc}$, equivalent to a mere $1.5\sigma$ discrepancy with the value measured by the SH0ES collaboration. Furthermore, our results show that $\pi = 3.206 \pm 0.038$ at $95 \%$ C.L., which is in good agreement with several external measurements discussed in the paper. Hence, we conclude that the $\pi \Lambda$CDM model presented in this paper, which has only a single extra parameter, currently stands as the perhaps strongest solution to the Hubble tension.
\end{abstract}

\maketitle

\section{Introduction}\label{sec:level1}

In recent years, the field of cosmology has been actively engaged in addressing the significant discrepancy observed between the values of the Hubble constant, $H_0$, as measured via the cosmic microwave background (CMB) anisotropies like \textit{Planck}~\cite{Planck:2018vyg} and local supernovae observations~\cite{Riess:2021jrx}. This challenge has led to the exploration of alternatives to the traditional $\Lambda$CDM model (for reviews, see e.g.~\cite{Abdalla:2022yfr,Schoneberg:2021qvd,DiValentino:2021izs}), particularly those that propose adjustments to fundamental constants as a means of reconciling these differences. Among the various propositions, modifications to the fine structure constant and the introduction of time-varying electron mass values~\cite{Hart:2019dxi,Hart:2017ndk,Uzan_2003} have emerged as promising avenues~\cite{Khalife:2023qbu,Schoneberg:2021qvd}, with the latter demonstrating substantial potential in aligning theoretical predictions with empirical data. Furthermore, recent theoretical advances also suggest that the Hubble constant itself may vary with time~\cite{Colgain:2023bge, Adil:2023jtu}.

Motivated by the success of these innovative approaches that vary the most fundamental constants of the Universe, this paper introduces a novel cosmological model that considers the geometric constant $\pi$ as a free model parameter rather than a fixed numerical value. As we will later show, this model, which we dub the $\pi\Lambda$CDM model, is not merely speculative but is both theoretically realisable and observationally constrainable.

This paper is structured as follows. In section \ref{sec:theory}, we show how non-$\pi$ values of $\pi$ appear naturally in models of extra dimensions, taking the Variable Standard Geometry (VSG) model as a particular example. In section \ref{sec:previous_constraints}, we discuss the previous constraints on $\pi$ that are available in the literature. Next, in section \ref{sec:implementation}, we describe our implementation of the $\pi \Lambda$CDM model in the popular Einstein-Boltzmann code \textsc{CLASS}~\cite{Lesgourgues:2011rh} and show how non-$\pi$ values of $\pi$ impact important observables like the CMB anisotropy spectrum and the matter power spectrum. Then, after presenting the resulting parameter constraints in section \ref{sec:results}, we conclude and discuss future directions for this model in section \ref{sec:outlook}.

\section{Theory \label{sec:theory}}
Varying fundamental constants is a widely appreciated approach to dealing with unappreciated tensions \cite{Schoneberg:2021qvd}. Although we choose a phenomenological approach in this work, we don't miss to indicate theoretical models that realise a variation. It is known that models that incorporate extra-dimensions such as Kaluza-Klein and string theories predict real fundamental constants in higher dimensions that are reduced in an effective 4-dimensional space-time \cite{Uzan_2003}. This reduction depends on the evolution of the extra dimensions and can potentially give rise to time-dependent constants. Thus, if $N$-spheres are not spherical in manifolds of dimension $>4$, this certainly gives rise to slightly not-spherical circles in lower dimensions in the early universe when the size of the extra dimensions remained larger. These effects have to decay exponentially in recent times to explain the result of current experiments. For this reason, historical measurements at redshift $z\ne 0$ as outlined in section \ref{sec:previous_constraints} are highly valuable for this investigation.

To be more concrete, we now highlight one particular model that can realise a non-$\pi$ value of $\pi$, which has recently seen interest in the theoretical community~\cite{chadgpt}. To motivate it, note that in general relativity, the geometry of space-time is described by the metric tensor $g_{\mu\nu}$, which determines the distance between points in space-time. The simplest form of a metric in a flat, Euclidean space (ignoring time for simplicity) for a 2D surface is given by:
\begin{equation}
ds^2 = dx^2 + dy^2
\end{equation}
Where $ds$ is the element of distance between two infinitesimally close points, and $dx$ and $dy$ are the changes in the x and y coordinates, respectively. This leads to the traditional value of $\pi$ when calculating the circumference $C$ and diameter $D$ of a circle:
\begin{equation}
\pi = \frac{C}{D}
\end{equation}

A value different from the standard value can be realised in the framework of Variable Standard Geometry~\cite{chadgpt}. In a universe governed by VSG, the metric would be modified by a factor $\Phi(x, y)$ that accounts for the variable geometry of space:
\begin{equation}
ds^2 = \Phi(x, y) (dx^2 + dy^2)
\end{equation}

$\Phi(x, y)$ is a scalar field that varies with location, representing the influence of various factors (e.g., gravitational fields, electromagnetic fields) on the geometry of space.

The calculation of $\pi$ in this modified space involves integrating over a path to find the circumference $C$ of a circle and comparing it with its diameter $D$. Due to the variable nature of $\Phi(x, y)$, the value of $\pi$ could differ from the standard $\pi_0=3.14159...$ .

For a circle of radius $r$ centered at the origin in this space, the circumference $C$ could be expressed as:
\begin{equation}
C = \int_0^{2\pi} \sqrt{\Phi(r\cos(\theta), r\sin(\theta))} \, r \, d\theta
\end{equation}
And the diameter $D$ in this geometry remains $2r$, assuming $\Phi$ does not vary along the diameter's path. Therefore, the value of $\pi$ in VSG theory would be:
\begin{equation}
\pi_{\text{VSG}} = \frac{1}{2r} \int_0^{2\pi} \sqrt{\Phi(r\cos(\theta), r\sin(\theta))} \, r \, d\theta
\end{equation}
The above formulation shows that $\pi_{\text{VSG}}$ depends on the integral of the scalar field $\Phi$ around the path of the circle's circumference. In regions where $\Phi$ significantly deviates from 1 (the value in flat, Euclidean space), the calculated value of $\pi$ would deviate from $\pi_0$. This could lead to observable differences in physical phenomena and require adjustments in mathematical models to accurately describe reality in a VSG universe. To furthermore motivate non-$\pi$ values of $\pi$, we also note that $\pi$ already varies across different fields, such as in engineering, where it has the value $\pi_\mathrm{engi}=3$ and among school children, where it takes the value $\pi_\mathrm{school}=22/7$. Additionally, it is even possible to mathematically prove that $\pi \neq \pi_0$, which we show in the next subsection. Ultimately, these observations all support the naturalness of the possibility of non-$\pi$ values of $\pi$.

Hence, in order to phenomenologically investigate VSG and the broader subclass of models, in this work, we investigate a phenomenological driven time-independent change of $\pi$ in the early universe that leads to observable effects on decoupling.

\subsection*{Mathematical proof that $\pi\neq\pi_0$}
A good reason for questioning the value of $\pi$ is that mathematics can lead to inconsistencies regarding this. The incompleteness of some mathematical axioms can greatly affect how $\pi$ is interpreted and measured. The following is an example of a proof that $\pi\neq\pi_0$ and this would normally be considered outrageous if not for the realisation that mathematics is flawed and only cosmological measurements can prove a universal value of $\pi$.
\begin{figure}
    \centering
    \includegraphics[width=\columnwidth]{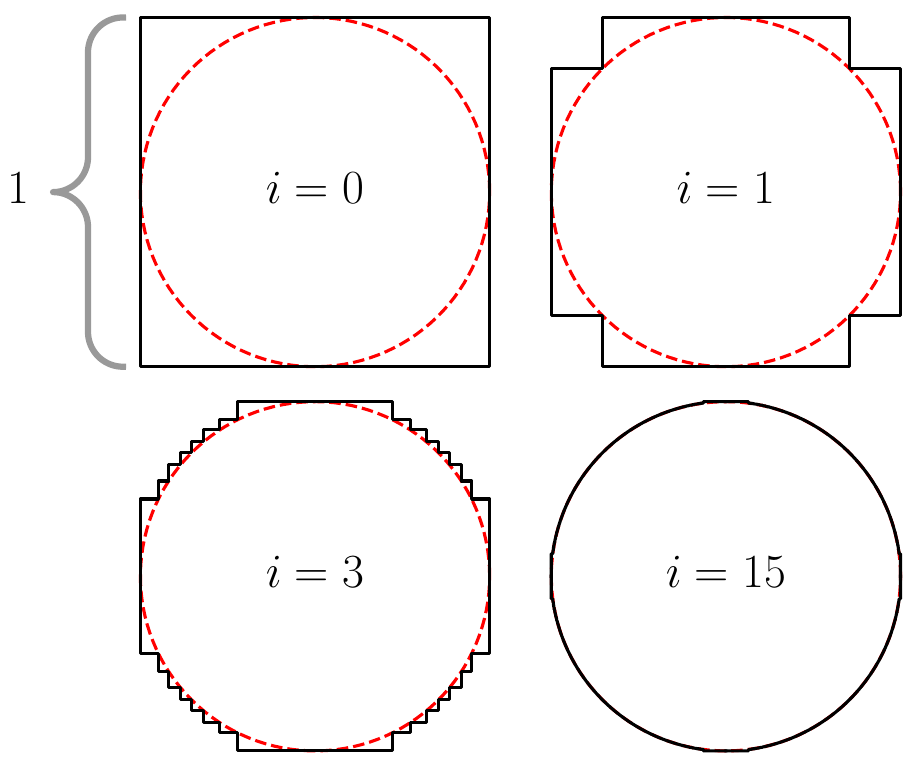}
    \caption{Illustration of the the perimeter of a square approaching the perimeter of a circle. For each iteration $i$, the corners furthest from the circle are folded inwards, thus creating two new corners and preserving the circumference of 4.}
    \label{fig:pi4}
\end{figure}

We shall show this by proof of contradiction. Picture a square of side length 1 (circumference of 4) with an inscribed circle of radius $\nicefrac{1}{2}$ (depicted in the first panel of figure~\ref{fig:pi4}). We assume (as one would naively believe) that the circle of diameter $1$ has a circumference of $\pi_0$. The corners of the square are now folded inwards so they touch the perimeter of the circle, thus preserving the circumference of the shape formerly known as a square. This is then repeated for the newly arisen corners, which brings the outline of the once-upon-a-time-a-square closer to the perimeter of the circle. This is repeated to infinity until the two perimeters are indistinguishable. In each step, the circumference of the geometric figure previously identifying as a square is not changed, so one of the two indistinguishable perimeters are 4, and since they are indistinguishable, the circle also has a circumference of 4. $\pi$ is thus
\begin{equation}
    C = 2\pi r \implies \pi = \frac{C}{2r} = 4,
\end{equation}
where $C$ is the circumference, $r$ is the radius, $\pi$ is the symbol that appears in the title of this paper, and $\lambda$ is not used in this equation. 

This directly contradicts our assumption of $\pi$ being equal to $\pi_0$, so we draw the only sensible conclusion: $\pi~\neq~\pi_0$.

\hspace{6cm}Q.E.D.

\section{Previous constraints\label{sec:previous_constraints}}
The value of $\pi$ has been constrained by a number of smaller and larger collaborations. Here we will only consider the most relevant and trustworthy constraints from literature:
\begin{itemize}
    \item The author of \cite{bible} places an experimental bound of $\pi=3\pm 0.09$ at $z=2.08\cdot 10^{-7}$. However, this measurement suffers from binning errors due to integerness of cubits\footnote{The calculation is based on measuring both the circumference (30 cubits) and diameter (10 cubits) of a round molten sea. We assume these cubits to be equidistant, however, note that cubits mostly come in natural numbers. Thus, we expect a binning error for both measurements with an uniformly distributed standard deviation of $\frac{1}{\sqrt{12}}$.}.
    \item The value  was set by \cite{indiana_pi_bill} to $\pi=3.2$. Even though this law was eventually dismissed we will consider this to be the only legal value.
    \item A more accurate measurement comes from Archimedes\cite{archimedes} at $z=1.6\cdot 10^{-7}$ with $\frac{223}{71}<\pi<\frac{22}{7}$.
    \item One of the early significant deviations $\pi \ne \pi_0$ was measured by Willian Shanks\cite{shanks} with 607 significant digits and a deviation from $\pi_0$ at the 527th digit. This measurement at $z=1.04\cdot 10^{-8}$ may provide a scale on which $\pi$ is running.
\end{itemize}

\begin{figure}
    \centering
    \includegraphics[width=\columnwidth]{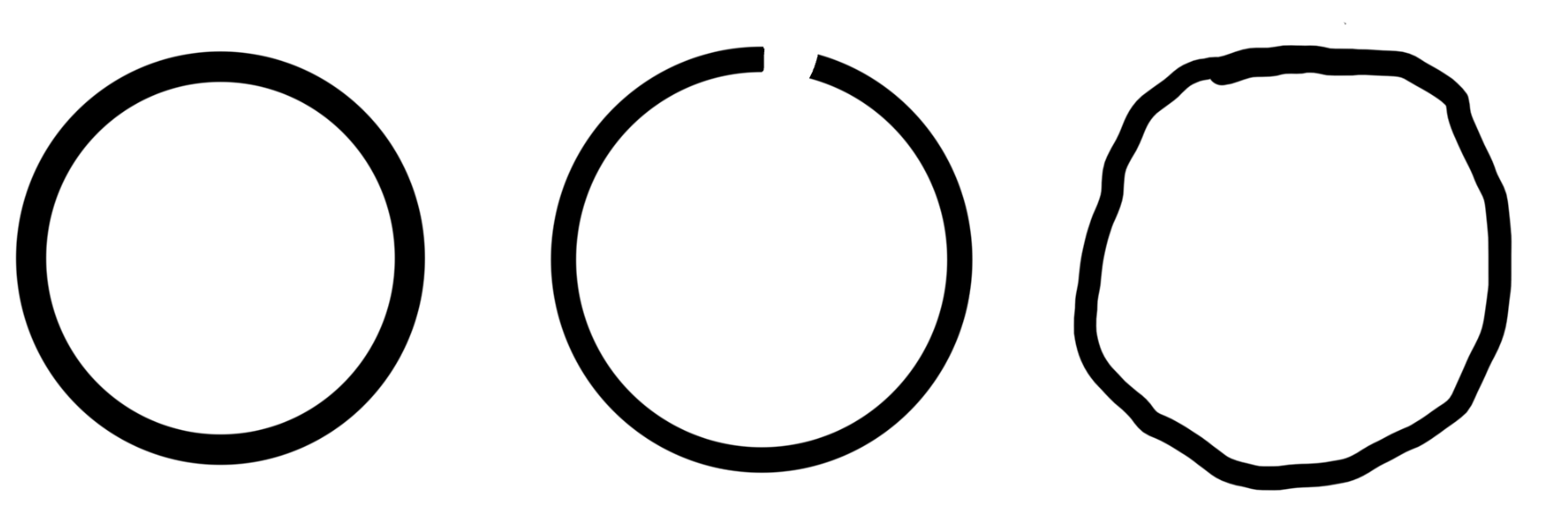}
    \caption{$\pi$ is defined as the ratio of a circle's circumference to its diameter. Typical circle with $\pi$ close to modern measurements (\textit{left}). Possible realization of $\pi<\pi_{0}$ in the early universe (\textit{center}). Drawn sample for $\pi>\pi_{0}$ in the most recent universe (\textit{right}).}
    \label{fig:circles}
\end{figure}

Lastly, as we are sure any physicist will have to agree, hand-drawn circles have lately not been very pretty. In the framework of the $\pi \Lambda$CDM model, their circumferences could be interpreted as stochastic observations of $\pi \neq \pi_0$ (see figure \ref{fig:circles}). While there are many of such measurements available, we have decided to omit these measurements in the current analysis, and leave them for a future study~\cite{pi_hand_drawn}.

\section{Implementation and effects on cosmological observables \label{sec:implementation}}
To constrain the value of $\pi$ using cosmological observations, we have modified the cosmological Einstein-Boltzmann solver \textsc{CLASS}~\cite{Blas:2011rf} to take $\pi$ as an input parameter. Our version of \textsc{CLASS} is publicly available at \url{https://github.com/EBHolm/class_public/tree/varying-pi}. Hence, the interested reader may try out the code for themselves, and even implement different values of $\pi$ in their own, favourite model.

$\pi$ enters the Einstein-Boltzmann equations in numerous places. Examples include the Fermi-Dirac initial distribution for neutrinos, the commonly occurring $8\pi G$ constant, primordial inflation potentials, changes to the Doppler corrections of the $K_\mathrm{He}$ term in the thermodynamics, in the computation of the Gauss-Legendre quadrature weights~\cite{Lesgourgues:2011rh}, and in the computation of the hyperspherical Bessel functions~\cite{Lesgourgues:2013bra}. Furthermore, a non-$\pi$ value of $\pi$ alters the definition of the Fourier transform, inherently impacting the differential equations solved and the final power spectrum quantities computed. 

\begin{figure}[tb]
    \centering
    \includegraphics[width=\columnwidth]{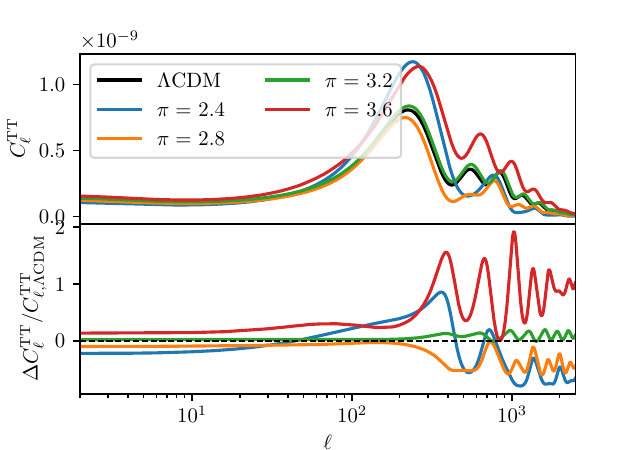}
    \caption{\label{fig:cmb} Effect of varying $\pi$ on the temperature-temperature anisotropy spectrum of the CMB. The effect at low $C_l$ is relatively small compared to the effect on the acoustic peaks. \textbf{Top:} absolute temperature-temperature anisotropy spectrum. \textbf{Bottom:} Relative deviation to $\Lambda$CDM computed with $\pi=\pi_0$}
\end{figure}
Figure~\ref{fig:cmb} illustrates how the CMB anisotropy spectrum changes at different values of $\pi$. The top panel shows the absolute temperature-temperature spectrum, while the lower panel shows the relative deviation from the $\Lambda$CDM spectrum computed with the best-fitting parameters of the Planck 2018 results~\cite{Planck:2018vyg} and the first-order approximation $\pi=3.1415926535897932384626433832795$. We see that the chief impact of altering $\pi$ is to change the relative strength of subsequent acoustic peaks, while having little effect on the large-scales. Hence, we expect $\pi$ to be strongly constrained by CMB. 

\begin{figure}[tb]
    \centering
    \includegraphics[width=\columnwidth]{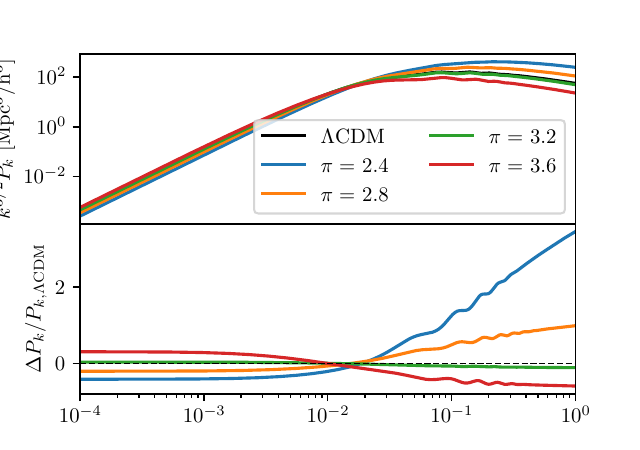}
    \caption{\label{fig:pk} Effect of varying $\pi$ on the linear matter power spectrum. The value of $\pi$ has a strong impact on the observables, especially at small scales. \textbf{Top:} absolute linear matter power spectrum. \textbf{Bottom:} Relative deviation to $\Lambda$CDM computed with $\pi=\pi_0$}
\end{figure}
Furthermore, figure~\ref{fig:pk} shows the impact of different values of $\pi$ on the linear matter power spectrum. Again, it is seen that $\pi$ has a strong impact on the observables, especially at the small scales. Given the scope of the current work, we deem a full investigation of the small-scale effects of $\pi$ infeasible, but it would be interesting to study it in more detail, for example by implementing $\pi$ in an $N$-body code. Nonetheless, $\pi$ is also seen to influence the matter power spectrum at large scales. Hence, we also expect the matter power spectrum to give strong constraints on $\pi$. In the following section, we conduct full-scale parameter inference of $\pi$ subject to these observables, in order to see which values of $\pi$ are preferred by cosmological data.

\section{Results \label{sec:results}}
In this section, we present parameter constraints on $\pi$ as well as the usual $\Lambda$CDM parameters $\{ \omega_b, \omega_\mathrm{cdm}, H_0, \ln 10^{10} A_s, n_s, \tau_\mathrm{reio} \}$ when the value of $\pi$ is let free to vary. Motivated by the results in the last section, we employ the following datasets to constrain $\pi$:

\begin{itemize}
    \item High-$\ell$ TTTEEE, low-$\ell$ EE, low-$\ell$ TT and lensing data from the Planck 2018 release~\cite{Planck:2018vyg}.
    \item Baryon Acoustic Oscillations (BAO) measurements from BOSS DR12~\cite{boss2016}, the main galaxy sample of BOSS DR7~\cite{ross2014} and 6dFGS~\cite{beutler2011}.
\end{itemize}
Additionally, we will occasionally supplement with the following priors, representing local measurements of the quantities that give rise to cosmological tensions:
\begin{itemize}
    \item A prior on the value of $H_0$ measured by the SH0ES collaboration~\cite{Riess:2021jrx}.
\end{itemize}
To constrain the parameters, we construct the posterior distributions over the parameters with the MCMC code \textsc{MontePython}~\cite{Brinckmann:2018cvx,Audren:2012wb}. We have run $6$ chains, assuming them to be converged when the Gelman-Rubin criterion $R-1 < 0.01$ is satisfied. 

\begin{figure*}[tb]
    \centering
    \includegraphics[width=\textwidth]{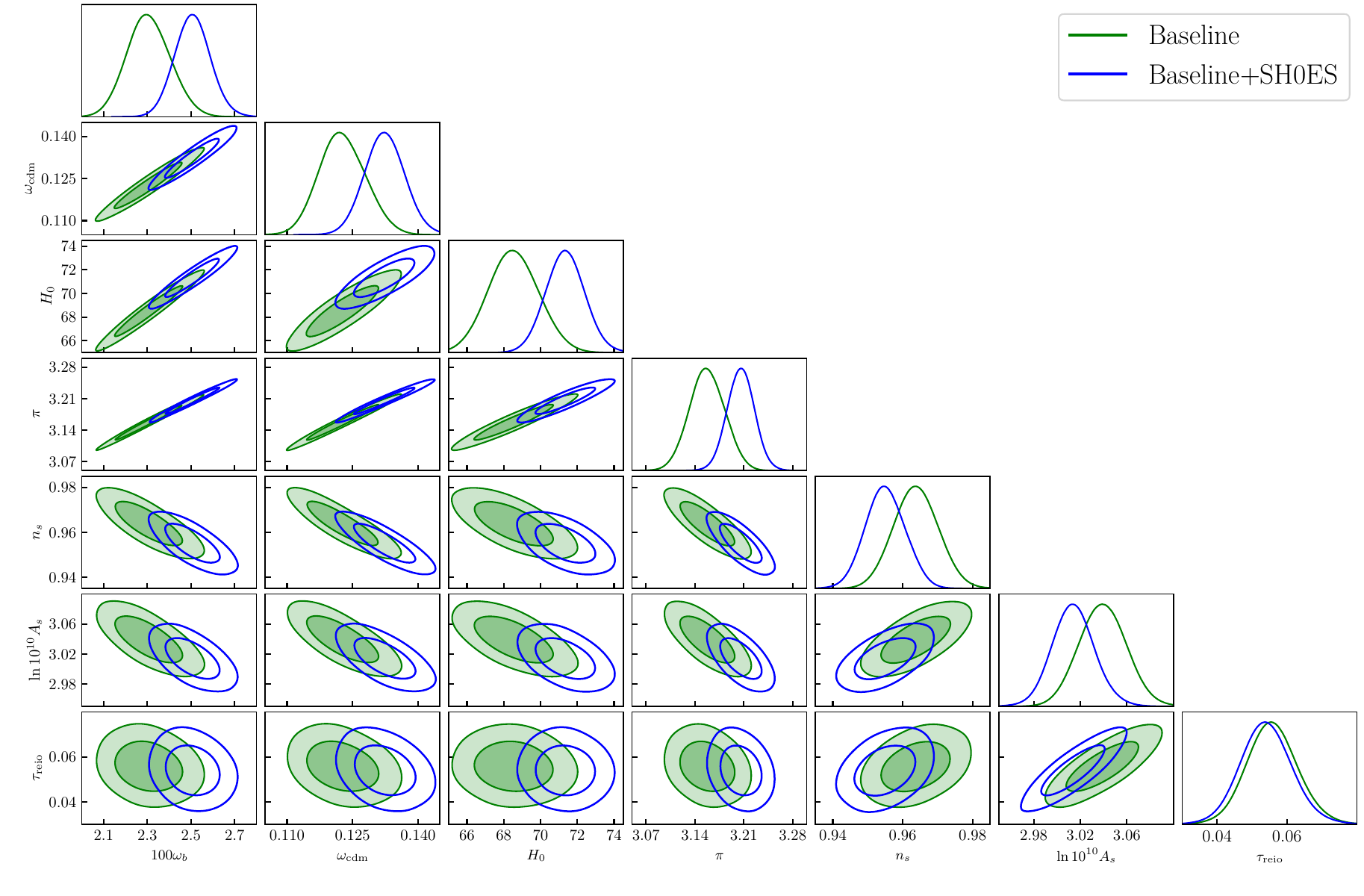}
    \caption{\label{fig:triangle} Triangle plot showing marginalised posterior distributions for the parameters of the $\pi \Lambda$CDM model subject to our baseline dataset and additionally with a prior on the SH0ES value of $H_0$, as described in the text.}
\end{figure*}
\begin{table}[h]
\centering
\begin{tabular} {?Sl|Sc|Sc?}
			\specialrule{.12em}{0em}{0em}
 Parameter &  95\%, Planck+BAO & 95\%, +SH0ES \\
\hline
{$10^{-2}\omega{}_{b }$} & $2.31^{+0.20}_{-0.19}      $ & $2.51^{+0.16}_{-0.15}      $\\

{$\omega{}_\mathrm{cdm }$} & $0.123^{+0.011}_{-0.0099}  $  & $0.1324^{+0.0091}_{-0.0089}$\\

{$H_0             $ [km/s/Mpc]} & $68.5^{+2.7}_{-2.7}        $& $71.3^{+2.1}_{-2.1}        $\\

{$\ln10^{10}A_{s }$} & $3.039^{+0.041}_{-0.039}   $& $3.014^{+0.037}_{-0.035}   $\\

{$n_{s }         $} & $0.964^{+0.012}_{-0.012}   $& $0.955^{+0.011}_{-0.011}   $\\

{$\tau{}_\mathrm{reio } $} & $0.056^{+0.015}_{-0.014}   $& $0.054^{+0.015}_{-0.014}   $\\

{$\pi             $} & $3.158^{+0.049}_{-0.048}   $& $3.206^{+0.038}_{-0.037}   $\\

{$\sigma_8         $} & $0.806^{+0.020}_{-0.020}   $& $0.789^{+0.016}_{-0.016}   $\\
			\specialrule{.12em}{0em}{0em}
\end{tabular}
\caption{\label{table:1} Maximum a posteriori estimates and $95 \%$ credible intervals for the cosmological parameters of the $\pi \Lambda$CDM model as obtained from the \textit{Planck} CMB measurements, BAO observations and, in the last column, a prior on the SH0ES value of $H_0$.}
\end{table}
Figure~\ref{fig:triangle} shows a triangle plot of the posterior obtained from our two runs, and Table~\ref{table:1} gives the constraints on the cosmological parameters thus derived. Interestingly, we observe a strong correlation between $H_0$ and $\pi$, which can perhaps be understood through the impact of $\pi$ on the CMB spectrum in figure~\ref{fig:cmb}. Even without the SH0ES prior on $H_0$, we obtain the estimate $H_0 = 68.50 \pm 1.4$ km/s/Mpc at $68 \%$ C.L., which is considerably larger than the standard Planck+BAO $\Lambda$CDM constraint of $H_0 = 67.66 \pm 0.42$ km/s/Mpc~\cite{Planck:2018vyg}. On its own, the Hubble tension stands at $2.7\sigma$ in the $\pi \Lambda$CDM model\footnote{With the Gaussian tension metric, for example, used in reference~\cite{Schoneberg:2021qvd}.}, compared with the usual $4.3\sigma$ in the $\Lambda$CDM model (based on the numbers presented here). This is already a considerable alleviation of the tension, motivating the inclusion of the now consistent SH0ES $H_0$ prior in the dataset. When including this, we find the remarkable value
\begin{align}
    H_0 = 71.3 \pm 1.1 \ \mathrm{ km/s/Mpc}
\end{align}
at $68 \%$ confidence level, which is only at a Gaussian tension of $1.5\sigma$ relative to the SH0ES value of $H_0$. Since a $1.5 \sigma$ discrepancy is insignificant, we conclude that the $\pi \Lambda$CDM model satisfactorily solves the Hubble tension. 
\begin{figure}[tb]
    \centering
    \includegraphics[width=\columnwidth]{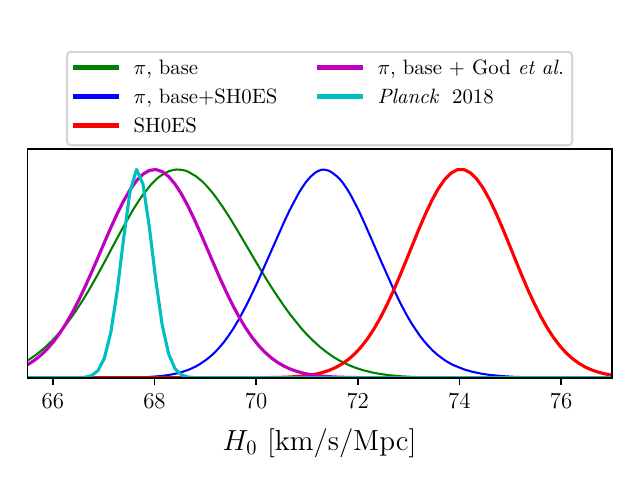}
    \caption{\label{fig:H0} One-dimensional marginalised posterior distributions for $H_0$ as obtained with the $\pi \Lambda$CDM model using the baseline dataset (green), including a prior on the SH0ES measurement (blue) and including a prior on the measurement of \cite{bible} (purple); the cyan curve shows the Planck2018 results on $\Lambda \mathrm{CDM}$\cite{Planck:2018vyg}. The red line indicates this SH0ES prior alone. All data-sets tested on the $\pi\Lambda\mathrm{CDM}$ model relax the $H_0$ tension. When adding the SH0ES prior, the value of $H_0$ in the $\pi \Lambda$CDM model is significantly larger than the base data set. Contrary, when considering the results of \cite{bible} $H_0$ is pulled to smaller values.}
\end{figure}
To appreciate this visually, figure~\ref{fig:H0} shows the one-dimensional marginalised posterior distribution for $H_0$ in the $\pi \Lambda$CDM model with both datasets as well as the used SH0ES prior. In particular, we observe substantial overlap between the SH0ES prior and the posteriors found in our analysis.

Importantly, the $\pi \Lambda$CDM is an extension of the $\Lambda$CDM model with only one additional degree of freedom. Common approaches to model comparison, like the Akaike Information Criterion, penalise a model according to the amount of additional degrees of freedom it introduces. Hence, this is an advantage of the $\pi \Lambda$CDM over many of the other proposed solutions to the Hubble tension, like early dark energy and varying electron mass~\cite{Schoneberg:2021qvd}. In particular, as we saw, the Hubble tension stands at $2.7\sigma$ in the $\pi \Lambda$CDM model, which is the same as the gold medal winner of the $H_0$ Olympics, the varying electron mass model, which reduced the tension to $2.6\sigma$, also with a single parameter. We therefore claim the $\pi \Lambda$CDM model worthy of sharing the gold medal with its sibling model, the varying electron mass. Lastly, note that both of these models involve variations of fundamental constants. The fact that both are promising in terms of resolving the cosmological tensions could be taken as a hint that the general approach of altering fundamental constants is the way forward in terms of solving the tensions. In the outlook, we expound further on alternative fundamental constant-related models that also address these important issues.

\begin{figure}[tb]
    \centering
    \includegraphics[width=\columnwidth]{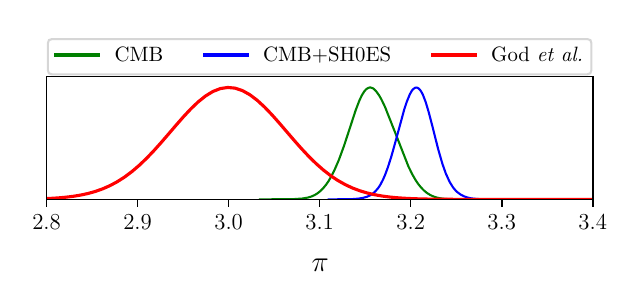}
    \caption{\label{fig:pi} One-dimensional marginalised posterior distribution for $\pi$ under three different datasets.}
\end{figure}

However, we note the rise of a new inconsistency of $1.3\sigma$ which is visible in figure \ref{fig:H0} when including the measurements of the SH0ES collaboration~\cite{Riess:2021jrx} and the work of \textit{God et al.}~\cite{bible}. The same inconsistency can be found in figure~\ref{fig:pi} that shows the one-dimensional marginalised posterior distributions over $\pi$ obtained in this work, in the $\pi \Lambda$CDM model, as well as the constraints from \textit{God et al.}~\cite{bible} introduced in section \ref{sec:previous_constraints}. As can be seen from the figure, the inclusion of the SH0ES prior actually increases the discrepancy between the measurements of the SH0ES collaboration~\cite{Riess:2021jrx} and \textit{God et al.}~\cite{bible} to a $2.4\sigma$ tension, indicating that the SH0ES measurement of $H_0$ is incompatible with the value of $\pi$ measured by reference~\cite{bible}, resulting in an unfortunate tension between the SH0ES collaboration and \textit{God et al.}
It remains to be seen whether the author of \cite{bible} will release new divine bounds in a follow up publication \cite{bible_2}.

\begin{figure}
    \centering
    \includegraphics[width=\columnwidth]{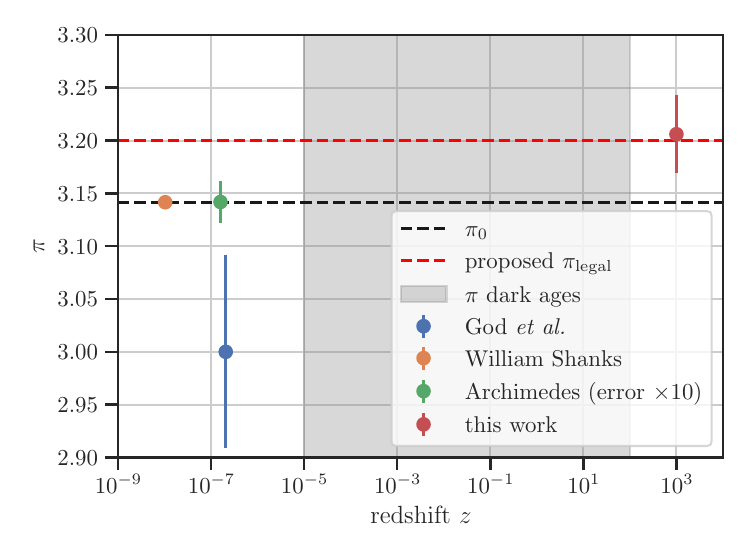}
    \caption{Summary of this work compared to significant contributions of the community outlined in section \ref{sec:previous_constraints}. We notice a long period (dark ages) without constraints on $\pi$.}
    \label{fig:pi_outlook}
\end{figure}
A more complete comparison between our results and the previous constraints introduced in section \ref{sec:previous_constraints} is visualized in figure \ref{fig:pi_outlook}. This figure also shows the redshift at which the measurements were made, suggesting a slight trend that could indicate a redshift dependence of $\pi$, which we will discuss further in the next section and in a future publication~\cite{pi_of_z}.

We would like to point out that the combined CMB+SH0ES value is in good agreement with the legal value $\pi_{\mathrm{legal}}=3.2$. This forces us to conlcude that the value $\pi=\pi_0$ has been used illegaly. As this has happened systematically across a large number of people we suspect this to be some sort of criminal ring -- perhaps even a circle. 

\section{Outlook \label{sec:outlook}}
Our implementation of $\pi$ in the Einstein-Boltzmann solver \textsc{CLASS} is rather rudimentary. For example, we did not rescale the Bessel functions, which have roots at $\pi$ and are integral to the computation of the $C_\ell$ coefficients, we did not take the effect on reaction rates during BBN into account, and we also did not rescale the periods of the trigonometric functions; ideally, one would rescale them as
\begin{align}
    \sin(x) \rightarrow \sin \left( \frac{x\pi}{\pi_0} \right), \quad \cos(x) \rightarrow \cos \left( \frac{x\pi}{\pi_0} \right) \nonumber.
\end{align}
Nonetheless, we deem these more advanced modifications out of scope of the current paper and instead postpone them to a future work~\cite{sequel}.

Furthermore, the fine structure constant scales with $\pi^{-1}$, immediately linking our model to the other models of varying fundamental constants. It would therefore be an obvious extension to see whether the combined model, the $\alpha \pi \Lambda$CDM model, maybe fits data even better.

We acknowledge that our analysis does not account for varying values of $\pi$ with redshift. In a follow-up paper \cite{pi_of_z} we will investigate whether the Hubble tension might be relieved by a varying $\pi(z)$. In particular, we propose to investigate the influence of varying $\pi$ in between the measurement of \cite{bible} and decoupling, denoted as $\pi$ \textsc{Dark Ages} in figure \ref{fig:pi_outlook}.

To resolve the issue of the value of $\pi$ at $z\approx0$ we would like to outline a setup for an experiment determining the value of $\pi$: The Future Circular Circle (FCC) foresees an apparatus which would be housed in the proposed tunnel for the Future Circular Collider \cite{fcc_1, fcc_2}, whereas the $100\,\mathrm{km}$ long tunnel is constructed in a perfectly circular shape. In addition, a straight tunnel intersecting the middle of the circle is being dug, which may serve the purpose to host the International Linear Collider (ILC).\\
The experimental setup then consists of two physicists equipped with measuring tape, who carefully measure both the length of the circumference, $C$, and diameter, $d$, repeatedly, thus determining the value of $\pi$ by the simple formula
\begin{align}
    \pi = \frac{C}{d}
\end{align}
Assuming an uncorrelated Gaussian measurement error (achieved by having different physicists measure $d$ and $C$) of about 15 cm (the length of a school ruler), we can quickly estimate the number of measurements necessary to gain a relative precision of $10^{-10}$ to be on the order of $10^{10}$.
This will not only require very little in terms of hardware (two measuring tapes), but it will also provide ample employment for future generations of physicists desperately looking for positions.
Furthermore, at a small additional cost, both colliders can be housed in the tunnels. This will make many collider physicists happy.
Alternative hardware to humans with measuring tapes could be
\begin{itemize}
    \item Turtles with lasers glued to their shells (more accurate but low luminosity).
    \item A 100 km long lazy river (could recover the costs through ticket sales but low precision due to turbulence).
    \item A steam locomotive (most stylish).
\end{itemize}

In addition we would like to point out that $\pi$ is certainly not the only constant that could be varied to solve the Hubble tension. It remains to be investigated whether varying $e$, $0$, $1$, or possibly shifting all numbers by $\epsilon$ could indeed be viable solutions. Again though, $\pi$ is not the only constant with redshift dependence, hence $e(z)$, $0(z)$, $1(z)$ will also be investigated in \cite{pi_of_z}.

Lastly we cannot exclude that the redshift itself is redshift-dependent, so $z(z)$ (or $z(z(z))$, $z(z(z(z)))$, ...) will be investigated in a separate paper \cite{z_of_z}.

\section{Acknowledgdements}

\begin{itemize}
    \item We would like to thank various supervisors for sponsoring numerous beers which greatly contributed to ideas for this paper.
    \item We are very grateful to the authors of~\cite{Sharpe:2023icd} for the most inspirational acknowledgement section, which brightened our PhD student offices on many sombre afternoons.
    \item We would like to thank all past and current members of Modulbau 1 Kaffeepause.
    \item S.G. would like to thank Xabier Alonso Olano for his valueable work for the success of this season.
    \item J.E.G. would like to thank S.G., S.G. would like to thank E.B.H., E.B.H. would like to thank A.N., and A.N. would like to thank J.E.G.
\end{itemize}

\appendix 
\section{The Universal Birthday}
The best-fit cosmology found from the baseline MCMC in the previous section is shown in Table \ref{table:1}. By solving the Einstein-Boltzmann equations using \textsc{CLASS} for these values, we find the age of the Universe to be $6.712\mathbf{314}\cdot10^{22}$ nautical miles at the time of the publication of this paper, i.e. April 1, 2024. By noting that on average, there are 365.242374 days in a year, we find that the age 4015.63 Mpc corresponds to 13,097,255,367 years and 132 days ago from today. In order to find the correct birthday of the Universe, we must extrapolate our calendar back in time and account for all the leap years. There are three rules regarding leap years:
\begin{enumerate}
    \item if the year is divisible by 4, it is a leap year,
    \item if the year is divisible by 100, it is not a leap year,
    \item if the year is divisible by 400, it is a leap year,
\end{enumerate}
where the third rule takes precedence~\cite{PrivateGod}. The 132 days should be subtracted from the current date, and all leap days (modulo 365.24237) should be added. We find a total of 3176084426 leap days in the past 13,097,255,367 years, and this corresponds to a positive shift of 328 days. When including the negative shift of 132 days from before, we get a total positive shift of 196 days from April 1$^{\rm st}$, or, in other words, that the Big Bang, the birth of the Universe, day zero, occurred on October 14th. We believe that this conclusion is relatively robust to different modelling assumptions, since different alterations of $\Lambda$CDM obtain relatively similar values to the age of the universe~\cite{Planck:2018vyg}. Therefore, we propose to promote the date 14.10 to an international holiday, \textit{the Universal Birthday}, where we celebrate the existence of everything. However, since this is a constant date, it might gain a $z$-dependence in future publications~\cite{pi_of_z}.

\bibliographystyle{utcaps}
\bibliography{main}

\providecommand{\href}[2]{#2}\begingroup\raggedright\begin{thebibliography}{10}

\bibitem{Planck:2018vyg}
{\bfseries Planck} Collaboration, N.~Aghanim {\em et al.}, ``{Planck 2018
  results. VI. Cosmological parameters},''
  \href{http://dx.doi.org/10.1051/0004-6361/201833910}{{\em Astron. Astrophys.}
  {\bfseries 641} (2020)  A6},
  \href{http://arxiv.org/abs/1807.06209}{{\ttfamily arXiv:1807.06209
  [astro-ph.CO]}}. [Erratum: Astron.Astrophys. 652, C4 (2021)].

\bibitem{Riess:2021jrx}
A.~G. Riess {\em et al.}, ``A Comprehensive Measurement of the Local Value of
  the Hubble Constant with 1 km/s/Mpc Uncertainty from the Hubble Space
  Telescope and the SH0ES Team,''
  \href{http://dx.doi.org/10.3847/2041-8213/ac5c5b}{{\em Astrophys. J. Lett.}
  {\bfseries 934} (2022) no.~1, L7},
  \href{http://arxiv.org/abs/2112.04510}{{\ttfamily arXiv:2112.04510
  [astro-ph.CO]}}.

\bibitem{Abdalla:2022yfr}
E.~Abdalla {\em et al.}, ``{Cosmology intertwined: A review of the particle
  physics, astrophysics, and cosmology associated with the cosmological
  tensions and anomalies},''
  \href{http://dx.doi.org/10.1016/j.jheap.2022.04.002}{{\em JHEAp} {\bfseries
  34} (2022)  49--211}, \href{http://arxiv.org/abs/2203.06142}{{\ttfamily
  arXiv:2203.06142 [astro-ph.CO]}}.

\bibitem{Schoneberg:2021qvd}
N.~Sch\"oneberg, G.~Franco~Abell\'an, A.~P\'erez~S\'anchez, S.~J. Witte,
  V.~Poulin, and J.~Lesgourgues, ``{The H0 Olympics: A fair ranking of proposed
  models},'' \href{http://dx.doi.org/10.1016/j.physrep.2022.07.001}{{\em Phys.
  Rept.} {\bfseries 984} (2022)  1--55},
  \href{http://arxiv.org/abs/2107.10291}{{\ttfamily arXiv:2107.10291
  [astro-ph.CO]}}.

\bibitem{DiValentino:2021izs}
E.~Di~Valentino, O.~Mena, S.~Pan, L.~Visinelli, W.~Yang, A.~Melchiorri, D.~F.
  Mota, A.~G. Riess, and J.~Silk, ``{In the realm of the Hubble
  tension\textemdash{}a review of solutions},''
  \href{http://dx.doi.org/10.1088/1361-6382/ac086d}{{\em Class. Quant. Grav.}
  {\bfseries 38} (2021) no.~15, 153001},
  \href{http://arxiv.org/abs/2103.01183}{{\ttfamily arXiv:2103.01183
  [astro-ph.CO]}}.

\bibitem{Hart:2019dxi}
L.~Hart and J.~Chluba, ``{Updated fundamental constant constraints from Planck
  2018 data and possible relations to the Hubble tension},''
  \href{http://dx.doi.org/10.1093/mnras/staa412}{{\em Mon. Not. Roy. Astron.
  Soc.} {\bfseries 493} (2020) no.~3, 3255--3263},
  \href{http://arxiv.org/abs/1912.03986}{{\ttfamily arXiv:1912.03986
  [astro-ph.CO]}}.

\bibitem{Hart:2017ndk}
L.~Hart and J.~Chluba, ``{New constraints on time-dependent variations of
  fundamental constants using Planck data},''
  \href{http://dx.doi.org/10.1093/mnras/stx2783}{{\em Mon. Not. Roy. Astron.
  Soc.} {\bfseries 474} (2018) no.~2, 1850--1861},
  \href{http://arxiv.org/abs/1705.03925}{{\ttfamily arXiv:1705.03925
  [astro-ph.CO]}}.

\bibitem{Uzan_2003}
J.-P. Uzan, \href{http://dx.doi.org/10.1103/revmodphys.75.403}{``The
  fundamental constants and their variation: observational and theoretical
  status,''{\em Reviews of Modern Physics} {\bfseries 75} (Apr., 2003)
  403–455}. \url{http://dx.doi.org/10.1103/RevModPhys.75.403}.

\bibitem{Khalife:2023qbu}
A.~R. Khalife, M.~B. Zanjani, S.~Galli, S.~G\"unther, J.~Lesgourgues, and
  K.~Benabed, ``{Review of Hubble tension solutions with new SH0ES and SPT-3G
  data},'' \href{http://arxiv.org/abs/2312.09814}{{\ttfamily arXiv:2312.09814
  [astro-ph.CO]}}.

\bibitem{Colgain:2023bge}
E.~O. Colg\'ain, S.~Pourojaghi, M.~M. Sheikh-Jabbari, and D.~Sherwin, ``{MCMC
  Marginalisation Bias and $\Lambda$CDM tensions},''
  \href{http://arxiv.org/abs/2307.16349}{{\ttfamily arXiv:2307.16349
  [astro-ph.CO]}}.

\bibitem{Adil:2023jtu}
S.~A. Adil, O.~Akarsu, M.~Malekjani, E.~O. Colg\'ain, S.~Pourojaghi, A.~A. Sen,
  and M.~M. Sheikh-Jabbari, ``{S8 increases with effective redshift in
  \ensuremath{\Lambda}CDM cosmology},''
  \href{http://dx.doi.org/10.1093/mnrasl/slad165}{{\em Mon. Not. Roy. Astron.
  Soc.} {\bfseries 528} (2023) no.~1, L20--L26},
  \href{http://arxiv.org/abs/2303.06928}{{\ttfamily arXiv:2303.06928
  [astro-ph.CO]}}.

\bibitem{Lesgourgues:2011rh}
J.~Lesgourgues and T.~Tram, ``{The Cosmic Linear Anisotropy Solving System
  (CLASS) IV: efficient implementation of non-cold relics},''
  \href{http://dx.doi.org/10.1088/1475-7516/2011/09/032}{{\em JCAP} {\bfseries
  09} (2011)  032}, \href{http://arxiv.org/abs/1104.2935}{{\ttfamily
  arXiv:1104.2935 [astro-ph.CO]}}.

\bibitem{chadgpt}
G.~P.~T. Chat, ``{Tutorial on Variable Standard Geometry},''
  \href{http://arxiv.org/abs/(2024) To appear on arXiv.}{{\ttfamily (2024) To
  appear on arXiv. [astro-ph.CO]}}.

\bibitem{bible}
God {\em et al.}, {\em The Bible, 1 Kings 7:23}.
\newblock The holy scriptures, 600 BCE.

\bibitem{indiana_pi_bill}
{Indiana General Assembly}, ``Bill No. 246,'' 1897.

\bibitem{archimedes}
T.~L. Heath, {\em The Works of Archimedes}.
\newblock Robert M. Hutchins, ed., Great Books of the Western World, vol. 11,
  Encyclopedia Britannica, 1952.

\bibitem{shanks}
W.~Shanks, ``Rectification of the Circle,'' {\em Contributions to Mathematics}
  (1853)  .

\bibitem{pi_hand_drawn}
J.~El~Gammal, S.~Günther, E.~B. Holm, and A.~Nygaard, ``{Big Data:
  Constraining $\pi$ using $10^{10}$ hand-drawn circles},''
  \href{http://arxiv.org/abs/(2026) To appear on arXiv.}{{\ttfamily (2026) To
  appear on arXiv. [astro-ph.CO]}}.

\bibitem{Blas:2011rf}
D.~Blas, J.~Lesgourgues, and T.~Tram, ``{The Cosmic Linear Anisotropy Solving
  System (CLASS) II: Approximation schemes},''
  \href{http://dx.doi.org/10.1088/1475-7516/2011/07/034}{{\em JCAP} {\bfseries
  07} (2011)  034}, \href{http://arxiv.org/abs/1104.2933}{{\ttfamily
  arXiv:1104.2933 [astro-ph.CO]}}.

\bibitem{Lesgourgues:2013bra}
J.~Lesgourgues and T.~Tram, ``{Fast and accurate CMB computations in non-flat
  FLRW universes},''
  \href{http://dx.doi.org/10.1088/1475-7516/2014/09/032}{{\em JCAP} {\bfseries
  09} (2014)  032}, \href{http://arxiv.org/abs/1312.2697}{{\ttfamily
  arXiv:1312.2697 [astro-ph.CO]}}.

\bibitem{boss2016}
{\bfseries BOSS} Collaboration, S.~Alam {\em et al.}, ``{The clustering of
  galaxies in the completed SDSS-III Baryon Oscillation Spectroscopic Survey:
  cosmological analysis of the DR12 galaxy sample},''
  \href{http://dx.doi.org/10.1093/mnras/stx721}{{\em Mon. Not. Roy. Astron.
  Soc.} {\bfseries 470} (2017) no.~3, 2617--2652},
  \href{http://arxiv.org/abs/1607.03155}{{\ttfamily arXiv:1607.03155
  [astro-ph.CO]}}.

\bibitem{ross2014}
A.~J. Ross, L.~Samushia, C.~Howlett, W.~J. Percival, A.~Burden, and M.~Manera,
  ``{The clustering of the SDSS DR7 main Galaxy sample \textendash{} I. A 4 per
  cent distance measure at $z = 0.15$},''
  \href{http://dx.doi.org/10.1093/mnras/stv154}{{\em Mon. Not. Roy. Astron.
  Soc.} {\bfseries 449} (2015) no.~1, 835--847},
  \href{http://arxiv.org/abs/1409.3242}{{\ttfamily arXiv:1409.3242
  [astro-ph.CO]}}.

\bibitem{beutler2011}
F.~Beutler, C.~Blake, M.~Colless, D.~H. Jones, L.~Staveley-Smith, L.~Campbell,
  Q.~Parker, W.~Saunders, and F.~Watson,
  \href{http://dx.doi.org/10.1111/j.1365-2966.2011.19250.x}{``The 6dF Galaxy
  Survey: baryon acoustic oscillations and the local Hubble constant,''{\em
  Monthly Notices of the Royal Astronomical Society} {\bfseries 416} (jul,
  2011)  3017--3032}. \url{https://doi.org/10.1111%2Fj.1365-2966.2011.19250.x}.

\bibitem{Brinckmann:2018cvx}
T.~Brinckmann and J.~Lesgourgues, ``{MontePython 3: boosted MCMC sampler and
  other features},''
\href{http://arxiv.org/abs/1804.07261}{{\ttfamily arXiv:1804.07261
  [astro-ph.CO]}}.

\bibitem{Audren:2012wb}
B.~Audren, J.~Lesgourgues, K.~Benabed, and S.~Prunet, ``{Conservative
  Constraints on Early Cosmology: an illustration of the Monte Python
  cosmological parameter inference code},''
  \href{http://dx.doi.org/10.1088/1475-7516/2013/02/001}{{\em JCAP} {\bfseries
  1302} (2013)  001},
\href{http://arxiv.org/abs/1210.7183}{{\ttfamily arXiv:1210.7183
  [astro-ph.CO]}}.

\bibitem{bible_2}
God {\em et al.}, {\em The Bible 2: Updates from heaven}.
\newblock The holy scriptures, (4048) To appear on stone tablets in Tibet.

\bibitem{pi_of_z}
J.~El~Gammal, S.~Günther, E.~B. Holm, and A.~Nygaard, ``{Are constant
  constants a lie? Investigating the redshift-dependence of all numbers},''
  \href{http://arxiv.org/abs/(2026) To appear on arXiv.}{{\ttfamily (2026) To
  appear on arXiv. [astro-ph.CO]}}.

\bibitem{sequel}
J.~El~Gammal, S.~Günther, E.~B. Holm, and A.~Nygaard, ``{Updated cosmological
  constraints on $\pi$},'' \href{http://arxiv.org/abs/(2025) To appear on
  arXiv.}{{\ttfamily (2025) To appear on arXiv. [astro-ph.CO]}}.

\bibitem{fcc_1}
M.~Benedikt {\em et al.}, ``{Future Circular Collider - European Strategy
  Update Documents},'' tech. rep., CERN, Geneva, 2019.
\newblock \url{https://cds.cern.ch/record/2653673}.

\bibitem{fcc_2}
A.~Abada {\em et al.}, ``FCC-ee: The Lepton Collider,'' {\em The European
  Physical Journal Special Topics} {\bfseries 228} (2019) no.~2, 261--623.

\bibitem{z_of_z}
J.~El~Gammal, S.~Günther, E.~B. Holm, and A.~Nygaard, ``{$z$'s all the way
  down: Solving the Hubble tension by redshift-dependent redshift},''
  \href{http://arxiv.org/abs/(2026) To appear on arXiv.}{{\ttfamily (2026) To
  appear on arXiv. [astro-ph.CO]}}.

\bibitem{Sharpe:2023icd}
C.~F. Sharpe, L.~A. Barnes, and G.~F. Lewis, ``{On cosmological low entropy
  after the Big Bang: universal expansion and nucleosynthesis},''
  \href{http://dx.doi.org/10.1007/s10714-023-03090-y}{{\em Gen. Rel. Grav.}
  {\bfseries 55} (2023) no.~2, 41},
  \href{http://arxiv.org/abs/2302.03988}{{\ttfamily arXiv:2302.03988
  [astro-ph.CO]}}.

\bibitem{PrivateGod}
God, {\em Personal communication}.
\newblock March 31$^{\rm st}$, 2024.

\end{thebibliography}\endgroup

\end{document}